\documentclass[journal,twocolumn,10pt]{IEEEtran}
\usepackage{textcomp}
\usepackage{stfloats}
\usepackage{url}
\usepackage{graphicx}
\usepackage{verbatim}
\usepackage{upgreek}
\usepackage{amssymb,amsmath}
\usepackage{color}
\usepackage{epstopdf}
\usepackage{bm}
\usepackage{cite}
\usepackage{array,color}
\usepackage{algorithm}
\usepackage{algpseudocode}
\usepackage{amsmath}
\usepackage{amsfonts}
\usepackage{amsthm}
\usepackage{graphics}
\usepackage{epsfig}
\usepackage{soul}
\usepackage{gensymb}
\soulregister\cite7
\soulregister\ref7
\usepackage{subfigure}
\usepackage{setspace}
\newtheorem{thm}{Theorem}
\newtheorem{rmk}{Remark}

\newtheorem{lemma}{Lemma}
\newtheorem{corollary}{Corollary}
\usepackage{titlesec} % 调整标题与正文之间的距离
\titlespacing{\section}{0pt}{*1.5}{*0.8} % {左边距} {前文距} {后文距} 
\titlespacing{\subsection}{0pt}{*1.5}{*0.8}

\begin{document}

\title{Doppler-Resilient LEO Satellite OFDM Transmission with Affine Frequency Domain Pilot}

\author{
	\IEEEauthorblockN{Shuntian Tang\IEEEauthorrefmark{1},
		Xiaomei Wu\IEEEauthorrefmark{2},
		Xinyi Wang\IEEEauthorrefmark{1},
		Le Zhao\IEEEauthorrefmark{1},
		Guang Yang\IEEEauthorrefmark{2},
		Zilong Liu\IEEEauthorrefmark{3},
		Fan Liu\IEEEauthorrefmark{4},
		 and
		Zesong Fei\IEEEauthorrefmark{1}}
		
	\IEEEauthorblockA{\IEEEauthorrefmark{1}School of Information and Electronics, Beijing Institute of Technology (BIT), Beijing 100081, China} \\
	\IEEEauthorblockA{\IEEEauthorrefmark{2}China Tower Corporation Limited, Beijing 100195, China} \\
	\IEEEauthorblockA{\IEEEauthorrefmark{3}School of Computer Science and Electronics Engineering, University of Essex, CO4 3SQ Colchester, U.K} \\
	\IEEEauthorblockA{\IEEEauthorrefmark{4}School of Information Science and Engineering, Southeast University, Nanjing, 211189, China} \\
	\IEEEauthorblockA{
		Email: 3120255634@bit.edu.cn}
}
%wuxm@chinatowercom.cn, wangxinyi@bit.edu.cn, tobin\_bit@icloud.com, \\
%yangguang891@chinatowercom.cn, zilong.liu@essex.ac.uk, and feizesong@bit.edu.cn
% The paper headers
%\markboth{Submitted to IEEE Transactions on Vehicular Technology,~Vol.~xx, No.~xx, xxx~2022}%
%{Shell \MakeLowercase{\textit{et al.}}: A Sample Article Using IEEEtran.cls for IEEE Journals}

%\IEEEpubid{0000--0000/00\$00.00~\copyright~2022 IEEE}
% Remember, if you use this you must call \IEEEpubidadjcol in the second
% column for its text to clear the IEEEpubid mark.
%\IEEEpubidadjcol

\maketitle

\begin{abstract}
Orthogonal frequency division multiplexing (OFDM) based low Earth orbit (LEO) satellite communication system suffers from severe Doppler shifts, while {the Doppler-resilient affine frequency-division multiplexing (AFDM) transmission suffers from significantly high processing complexity in data detection}. In this paper, we explore the channel estimation gain of affine frequency (AF) domain pilot to enhance the OFDM transmission under high mobility. Specifically, we propose a novel AF domain pilot embedding scheme for satellite-ground downlink OFDM systems for capturing the channel characteristics. By exploiting the autoregressive (AR) property of adjacent channels, a long short-term memory (LSTM) based predictor is designed to replace conventional interpolation operation in OFDM channel estimation. Simulation results show that the proposed transmission scheme significantly outperforms conventional OFDM scheme in terms of bit error rate (BER) under high Doppler scenarios, thus paving a new way for the design of next generation non-terrestrial network (NTN) communication systems.
\end{abstract}


\section{Introduction}

\IEEEPARstart{L}{ow} Earth orbit (LEO) satellite communication has attracted increasing research attention from both academia and industry in recent years~\cite{Kodheli2021COMST}. Particularly, the inclusion of non-terrestrial network (NTN) components in the third generation partnership project (3GPP) release-17 has further accelerated research and standardization efforts in this area~\cite{3gpp.38.821}. Compared with conventional terrestrial networks (TNs), NTNs enable seamless global connectivity independent of geographical or topographical constraints, making them a key enabler for ubiquitous communications. 
%~\cite{Yue2023COMST}

One major challenges in LEO satellite communication lies in the significant Doppler effect caused by its extremely high mobility. For instance, at a carrier frequency of 2 GHz, the maximum Doppler shift can reach up to 48 kHz~\cite{3gpp.38.821}. However, current wireless communication systems predominantly employ orthogonal frequency division multiplexing (OFDM) technology, which is highly sensitive to high Doppler. The widely employed solution is to leverage ephemeris information to pre-compensate Doppler shifts, with the residual Doppler effect estimated by pilots~\cite{3gpp.38.821}. However, due to inevitable inaccuracies in ephemeris data and the limited precision of pilot-based estimators, it remains challenging to ensure the communication reliability~\cite{He2025JSAC}.

As a promising technique against Doppler effects, affine frequency division multiplexing (AFDM) technology exhibits superior performance in rapidly time-varying channels~\cite{Rou2024MSP}. Specifically, AFDM employs the inverse discrete affine Fourier transform (IDAFT) with tunable parameters to effectively separate multipath components in the affine frequency (AF) domain, thereby achieving full diversity gains. In~\cite{Tang2025ICCC}, the authors leveraged a sparse Bayesian learning framework to design an AFDM-used channel estimator, demonstrating strong performance under high-mobility conditions. Furthermore, {for multiple-input–multiple-output AFDM (MIMO-AFDM) systems, \cite{Luo2025TWC} presented a joint data detection and decoding receiver with sparse graph theory. In \cite{Sui2025TWC}, the authors investigated a generalized spatial modulation-aided AFDM (GSM-AFDM) scheme, demonstrating superior BER performance compared with conventional AFDM. Moreover, \cite{Luo2024TWC} addressed massive connectivity challenges in high-mobility environments using AFDM, while \cite{sui2025multi} and~\cite{yin2025arxivAffine} explored AFDM’s potential in integrated sensing and communication (ISAC), and satellite communication scenarios, respectively.} 

Although the aforementioned works have validated that AFDM enables more accurate acquisition of channel state information (CSI), it should be noted that to obtain the full diversity gain, complex data detection is required, resulting in high processing latency and complexity. 
Furthermore, compared with OFDM, AFDM also suffers from less flexible multiple access capability, thus hindering its wide applicability. 

In this paper, we explore the strong channel estimation capability of AF-domain pilot under high mobility to enhance OFDM transmission in LEO satellite communications. We firstly study the relationship between the frequency domain and AF domain representations of the channel, and propose a novel AF-domain pilot assisted OFDM channel estimation scheme that fully captures channel characteristics. To address the rapid CSI variation under high mobility, we exploit the autoregressive (AR) property of the channel from the perspective of difference equations and design a long short-term memory (LSTM) network leveraging this property. Simulation results demonstrate that the proposed transmission scheme achieves significant improvements compared to conventional OFDM scheme under high Doppler conditions.

\section{AF-Domain Pilot Aided OFDM Transmission Scheme} \label{II}
In this section, we first present the LEO satellite channel model. Subsequently, we propose an AF-domain pilot embedding scheme for improving OFDM channel estimation by leveraging the equivalence between AF and time-frequency (TF) domain CSI.
\subsection{Satellite Communication Channel Model}
% We first introduce the doubly selective channel model in the time domain, frequency domain, and affine frequency domain to establish the transformation relationships among them. This lays the foundation for obtaining the channel state information (CSI) of OFDM systems in the affine frequency domain.
The LEO satellite-ground communication channel is typically considered to be time and frequency doubly selective, with its time-delay-domain channel impulse response (CIR) expressed as~\cite{Bemani2023TWC}
\begin{equation} \label{CIR}
	h_{\text{TD}}(t,\tau)=\sum_{i=0}^{P-1}h_ie^{-j2\pi f_it}\delta(\tau-\tau_i),
\end{equation}
where $P$ is the number of propagation paths, $h_i$, $l_i$, and $f_i$ denote the complex channel gain, the transmission time delay, and the Doppler shift of the $i$-th path, respectively.
% where $P$ is the number of propagation paths, and $h_i$, $l_i$, and $f_i$ denote the complex channel gain, the normalized time delay (with respect to the sampling period), and the Doppler shift (in digital frequency) of the $i$-th path, respectively.

For an OFDM system with $N$ subcarriers and subcarrier spacing $\Delta f$, the frequency-domain channel matrix associated with $k$-th OFDM symbol, based on (\ref{CIR}), can be derived as~\cite{Ma2025WCL}
\begin{equation} \label{CFR}
	\mathbf{H}_{\text{FD},k}=\sum_{i=0}^{P-1}\tilde{h}_{i,k}\mathbf{F}\mathbf{\Delta}_{f_i}\mathbf{\Pi}^{l_i}\mathbf{F}^H,
\end{equation}
where $\mathbf{F}$ is the normalized $N$-point DFT matrix, $\mathbf{\Delta}_{f_i} = \text{diag}\left([1,e^{-j2\pi f_i},\cdots,e^{-j2\pi f_i\left(N-1\right)}]\right)$ is the Doppler shift matrix, $\mathbf{\Pi}^{l_i}$ denotes the cyclic permutation matrix corresponding to the normalized delay $l_i$ with respect to the sampling period $t_s = 1/f_s=1/(N \Delta f)$, and $\tilde{h}_{i,k}$ denotes the time-varying channel gain given by 
\begin{equation} \label{htilde}
	\tilde{h}_{i,k}={h}_{i}e^{-j2\pi f_c/f_s l_i}e^{-j2\pi \nu_i[(k+1)L+Nk]/N},
\end{equation}
where $f_c$ denotes the carrier frequency, $\nu_i = N f_i$ denotes the normalized Doppler shift, and $L$ denotes the length of cyclic prefix (CP).

Let $\nu_i = \alpha_i + a_i$ characterize the Doppler effect, where $\alpha_i \in [-\alpha_{\max}, \alpha_{\max}]$ is the integer component while $a_i \in (-0.5, 0.5]$. Under high-mobility conditions, fractional Doppler $a_i$ destroys subcarrier orthogonality and spreads energy across adjacent subcarriers, making single-tone pilots unable to capture the true channel response. Meanwhile, the channel varies rapidly from symbol to symbol; hence, interpolation between pilot symbols becomes unreliable. As a result, conventional OFDM channel estimation suffers severe degradation in the LEO satellite system.

\subsection{Doppler-Resilient OFDM Transmission with AF-Domain Pilot}
Unlike conventional OFDM estimation schemes, AF-domain channel estimation explicitly estimates $\tilde{h}_{i,k}$, $l_i$, and $\nu_i$ for each path and then reconstructs the full channel matrix~\cite{Bemani2023TWC}. This sensing-like procedure yields a more complete and structurally faithful representation of the channel. Motivated by this, we propose an AF-domain channel estimation enhanced Doppler-resilient OFDM transmission framework as follows.

% For a transmitted frequency domain OFDM symbol $\mathbf{x} \in \mathbb{C}^{N \times 1}$, the received frequency domain symbol after passing through the NTN channel is given by
%\begin{equation} \label{IO-OFDM}
%	\mathbf{y}=\mathbf{H}_{\text{FD}}\mathbf{x}+\mathbf{w},
%\end{equation}
%where $\mathbf{w} \sim \mathcal{CN}(0, \sigma^2\mathbf{I}_N)$ represents additive white Gaussian noise (AWGN) at the receiver.

Before presenting the proposed scheme, we first analyze the doubly selective channel matrix of (\ref{CIR}) in the AF domain with chirp parameters $c_1$ and $c_2$, which is given by~\cite{Bemani2023TWC}
\begin{equation} \label{CAFR}
	\mathbf{H}_{\text{AFD},k}=\sum_{i=1}^{P}\tilde{h}_{i,k}\mathbf{\Lambda}_{c_2}\mathbf{F}\mathbf{\Lambda}_{c_1}\mathbf{\Gamma}_i\mathbf{\Delta}_{f_i}\mathbf{\Pi}^{l_i}\mathbf{\Lambda}_{c_1}^H\mathbf{F}^H\mathbf{\Lambda}_{c_2}^H,
\end{equation}
where $\mathbf{\Lambda}_{c}=\text{diag}\left([1,e^{-j2\pi c},\cdots,e^{-j2\pi c\left(N-1\right)^2}]\right)$ is the chirp modulation matrix, and $\mathbf{\Gamma}_i$ is an $N \times N$ diagonal matrix defined as
\begin{equation}
	\mathbf{\Gamma}_{i}(n,n)= \begin{cases}e^{-j 2 \pi c_1\left(N^2-2 N\left(l_i-n\right)\right)}, & n<l_i, \\ 1, & n \geq l_i.\end{cases}
\end{equation}
It is worth noting that when $2Nc_1$ is an integer and $N$ is even, $\mathbf{\Gamma}_i = \mathbf{I}_N$.

To achieve full diversity gain, the chirp parameter $c_1$ should satisfy the following condition~\cite{Bemani2021ICC}
\begin{equation} \label{c1}
	c_1 = \frac{2(\alpha_{max}+k_v)+1}{2N},
\end{equation}
where $k_v$ is an integer to ensure that different multipath components are not overlapped as much as possible. 

In practical applications, $N$ is typically chosen as an even number to facilitate the use of the fast Fourier transform (FFT); hence, $2Nc_1$ is usually an integer. As a result, $\mathbf{\Gamma}_i$ simplifies to $\mathbf{I}_N$, and the channel matrix in the AF domain can be simplified as
\begin{equation} \label{CAFRs}
	\mathbf{H}_{\text{AFD},k}=\sum_{i=1}^{P}\tilde{h}_{i,k}\mathbf{\Lambda}_{c_2}\mathbf{F}\mathbf{\Lambda}_{c_1}\mathbf{\Delta}_{f_i}\mathbf{\Pi}^{l_i}\mathbf{\Lambda}_{c_1}^H\mathbf{F}^H\mathbf{\Lambda}_{c_2}^H.
\end{equation}

%For a transmitted AFDM symbol $\mathbf{x} \in \mathbb{C}^{N \times 1}$ in the frequency domain, the corresponding received symbol is given by
%\begin{equation} \label{IO-AFDM}
%	\mathbf{y}=\mathbf{H}_{\text{FD}}\mathbf{x}+\widetilde{\mathbf{w}},
%\end{equation}
%where $\widetilde{\mathbf{w}} \sim \mathcal{CN}(0, \sigma^2 \mathbf{I}_N)$ has the same statistics with ${\mathbf{w}}$.

\begin{figure*}[t]
	\centering
	\includegraphics[width=0.9\textwidth]{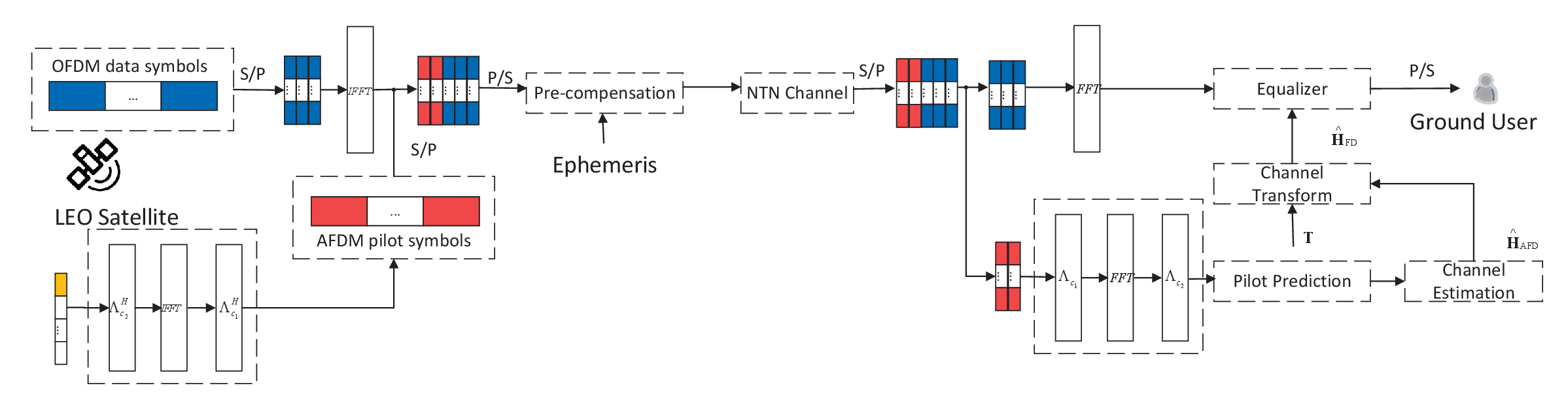}
	\caption{Block diagram of AF-domain pilot embedding enhanced LEO-OFDM system.}
	\label{fig:sys}
\end{figure*}

By defining the unitary transform matrix $\mathbf{T}=\mathbf{\Lambda}_{c_2}\mathbf{F}\mathbf{\Lambda}_{c_1}\mathbf{F}^H$, a relationship between $\mathbf{H}_{\text{FD},k}$ and $\mathbf{H}_{\text{AFD},k}$ can be established as
\begin{equation} \label{AO-relation}
	\mathbf{H}_{\text{FD},k}=\mathbf{T}^H\mathbf{H}_{\text{AFD},k}\mathbf{T}.
\end{equation}
This relationship implies that the TF domain CSI can be obtained from AF-domain CSI. 
%\begin{rmk}
%	For frequency selective channels with low mobility, the above channel estimation process by (\ref{AO-relation}) is more accurate based on the simulation results. A conventional approach based on subcarrier-wise estimation and interpolation in the frequency domain is typically sufficient. However, in doubly selective channels with high Doppler frequency, such frequency-domain methods become inaccurate, and the doubly selective channel is quasi-static in the affine frequency domain, which enables accurate channel estimation when adopting the AFDM-based estimation framework.
%\end{rmk}
Building upon the insight provided by (\ref{AO-relation}), the block diagram of AF-domain pilot enhanced LEO satellite OFDM systems is illustrated in Fig.~\ref{fig:sys} (shown at the top of the next page), where the 
conventional OFDM pilot symbols are replaced by AF-domain pilots which are generated by applying the IDAFT to a single-element pilot in the AF domain. Owing to the properties of the IDAFT, the resulting time-domain pilot symbols exhibit a constant modulus, achieving low peak-to-average power ratio (PAPR). These pilot sequence are placed sequentially at the beginning of each time slot, providing the basis for channel prediction at the receiver, which is detailed in the subsequent section. {Specifically, the transmitted time-domain signal per slot can be expressed as}
\begin{equation} \label{transig}
	{\mathbf{X}=[\mathbf{x}_{p,0},\ldots,\mathbf{x}_{p,L-1},\mathbf{x}_{d,0},\ldots,\mathbf{x}_{d,S-1}]^T \in \mathbb{C}^{N\times N_{sym}},}
\end{equation}
{where $\mathbf{x}_{p,i} = [1,\ldots,e^{j2\pi c_1(N-1)^2}]^T$ denotes the AF-domain pilot symbols and $\mathbf{x}_{d,i}$ denotes the data symbols. The pilot-to-data symbol number ratio (P/D ratio), defined as $L/S$, is used to quantify the transmission efficiency.}

\begin{rmk}
	Although ephemeris-based Doppler pre-compensation is supported at the transmitter side, its effectiveness is limited by the accuracy of ephemeris data~\cite{Sun2024ICCTIT}. According to~\cite{3gpp.38.821}, the residual Doppler shift can reach up to 1.05~ppm, corresponding to approximately 2100~Hz at a carrier frequency of 2~GHz. Therefore, additional mechanisms are required to mitigate the residual Doppler shift at the receiver.
\end{rmk}

The pilot sequence in the AF domain possesses an AR structure (as discussed in the next section), which enables the inference of subsequent received pilot symbols from only a subset of the observed pilots. This implies that for resource elements embedded with data symbols,  we can infer what the received symbol would have been if a pilot had been placed there, allowing us to estimate the corresponding CSI, effectively creating \textbf{virtual pilot symbols} in the inference process. Based on these inferred virtual pilots, channel estimation is then performed in the AF domain. Subsequently, the TF-domain channel matrix is reconstructed using the relation in (\ref{AO-relation}).

%\begin{rmk}
%	\textcolor{blue}{Although the obtained TF-domain channel matrix is two-dimensional, it remains diagonally dominant when the integer Doppler component is eliminated through Doppler pre-compensation. Such matrices can be inverted efficiently using low-complexity iterative algorithms, such as Jacobi or Gauss–Seidel methods, leading to substantially lower computational cost compared with equalization carried out directly in the AF domain.}
%\end{rmk}

\section{LSTM-based Virtual Received Pilots Prediction based Channel Prediction}

In this section, we investigate the design of the pilot prediction module shown in Fig.~\ref{fig:sys}. We first analyze the inherent AR structure of the received pilot sequence and then propose a LSTM neural network (NN) for pilot prediction.

\subsection{AR Property of Received Pilot Sequence}
To characterize the relationship among received pilot symbols, we assume that the transmitter sends the same pilot $\mathbf{x}_{\text{p}}$ on each symbol. After propagating through the $P$-path channel described in (\ref{CIR}), the $k$-th received pilot symbols in the AF domain can be expressed as
\begin{equation} \label{ypk}
	\mathbf{y}_{\text{p},k}=\mathbf{H}_{\text{AFD},k}\mathbf{x}_{\text{p}}=\sum_{i=0}^{P-1}\tilde{h}_{i,k}\mathbf{H}_{i}\mathbf{x}_{\text{p}},
\end{equation}
where $\mathbf{H}_{i}=\mathbf{\Lambda}_{c_2}\mathbf{F}\mathbf{\Lambda}_{c_1}\mathbf{\Delta}_{f_i}\mathbf{\Pi}^{l_i}\mathbf{\Lambda}_{c_1}^H\mathbf{F}^H\mathbf{\Lambda}_{c_2}^H$. Furthermore, the $n$-th element of $k$-th received pilot symbols can be expressed as
\begin{align}\label{ypkn}
	{y}_{\text{p},k,n}&=\sum_{i=0}^{P-1}\tilde{h}_{i,k}\mathbf{H}_{i}(n,:)\mathbf{x}_{\text{p}},\\ \notag
	%&=\sum_{i=0}^{P-1}\tilde{h}_{i,0}e^{-j2\pi \nu_i(N+L)k/N}\mathbf{H}_{i}(n,:)\mathbf{x}_{\text{p}},\\ \notag
	&\triangleq\sum_{i=0}^{P-1}c_{n,i}e^{-j\theta_ik},
\end{align}
where $\mathbf{H}_{i}(n,:)$ denotes the $n$-th row vector of $\mathbf{H}_{i}$, $c_{n,i}=\tilde{h}_{i,0}\mathbf{H}_{i}(n,:)\mathbf{x}_{\text{p}}$, and $\theta_i=2\pi \nu_i(N+L)/N$. 

As shown in (\ref{ypkn}), for any fixed index $n$, the sequence $\{{y}_{\text{p},k,n}\}(k=0,1,\ldots)$ is a linear combination of $P$ complex exponential terms $\{e^{-j\theta_ik}\}(i=0,1,\ldots,P-1)$. It is a well-established principle that any sequence formed as a finite sum of exponential terms satisfies a $P$-th order homogeneous linear difference equation with constant coefficients. In particular, the roots of the characteristic polynomial of this difference equation are $\{e^{-j\theta_i}\}(i=0,1,\ldots,P-1)$. Therefore, there exists a set of constant coefficients $\{\beta_0,\ldots,\beta_{P-1}\}$ such that
\begin{equation}\label{nDQ}
	y_{\text{p},k+P,n} + \beta_{P-1}y_{\text{p},k+P-1,n} + \cdots + \beta_0 y_{\text{p},k,n} = 0,\quad \forall k,
\end{equation}
which implies that future sample $y_{\text{p},k+P,n}$ can be predicted from the past $P$ samples $\{y_{\text{p},k,n},\dots,y_{\text{p},k+P-1,n}\}$.

It can be observed that each component of $\mathbf{y}_{\text{p},k}$ satisfies a similar difference equation, implying that a predictive relationship exists for each element. In particular, since the roots of the characteristic polynomials corresponding to these difference equations are identical, i.e. $\{e^{-j\theta_i}\}(i=0,1,\ldots,P-1)$, all components share the same set of coefficients. Consequently, the received pilot symbols sequence satisfies a unified $P$-th order vector difference equation:
\begin{equation}\label{vDQ}
	\mathbf{y}_{\text{p},k+P} + \beta_{P-1}\mathbf{y}_{\text{p},k+P-1} + \cdots + \beta_0 \mathbf{y}_{\text{p},k} = \mathbf{0},\quad \forall k,
\end{equation}
which can be further transformed as
\begin{equation}\label{yAR}
	\mathbf{y}_{\text{p},k+P}=\gamma_{0}\mathbf{y}_{\text{p},k} + \cdots + \gamma_{P-1} \mathbf{y}_{\text{p},k+P-1},\quad \forall k.
\end{equation}

Based on (\ref{yAR}), the sequence of received pilot symbols exhibits an AR property, providing the theoretical foundation for the aforementioned inference of virtual pilot symbols. Building upon this analysis, we present the following conclusion, formally stated as \textbf{Lemma~\ref{lemma1}}.
\begin{lemma} \label{lemma1}
Assume that the transmitter sends the same pilot symbol $\mathbf{x}_{\text{p}}$ on each symbol in the AF domain within the channel coherence time and bandwidth. The sequence of received pilot symbols $\{\mathbf{y}_{\text{p},k}\}$ in the AF domain possesses the $P$-order AR property described in (\ref{yAR}).
\end{lemma}
Similarly, \textbf{Lemma~\ref{lemma1}} can be extended to both TF domain and delay–Doppler domain. Moreover, the AR property is not limited to the received pilot sequence; the corresponding channel matrix also exhibits a similar AR structure. In essence, the $P$-th order AR behavior of the received pilot sequence arises from the inherent AR property of the channel matrix itself, leading to the following \textbf{Theorem~\ref{theorem1}}.
\begin{thm} \label{theorem1}
	Within the channel coherence time and bandwidth, the channel matrix $\{\mathbf{H}_{k}\}$ possesses the $P$-order AR property, which can be expressed as 
	\begin{equation}\label{HAR}
		\mathbf{H}_{k+P}=\gamma_{0}\mathbf{H}_{k} + \cdots + \gamma_{P-1} \mathbf{H}_{k+P-1},\quad \forall k.
	\end{equation}
\end{thm}

In static or low-Doppler scenarios, interpolation between observed pilot symbols can approximately capture the relationship described in (\ref{HAR}). However, as the Doppler shift increases, interpolation becomes ineffective, leading to significant performance degradation. Motivated by (\ref{HAR}), the future CSI can instead be predicted using the most recent $P$ channel matrices. Specifically, the AR coefficients $\gamma_{k}$ $(k = 0, \ldots, P-1)$ are first estimated from the previous $P+1$ received pilot symbols using the least-squares (LS) method, expressed as
\begin{equation}\label{LS}
 	\bm{\gamma}=\left(\mathbf{Y}\mathbf{Y}^H\right)^{-1}\mathbf{Y}^H\mathbf{y}_{k+P},
\end{equation}
where $\bm{\gamma} = [\gamma_0, \ldots, \gamma_{P-1}]^T$, and $\mathbf{Y} = [\mathbf{y}_{k}, \ldots, \mathbf{y}_{k+P-1}]$. Once $\bm{\gamma}$ is obtained, the subsequent channel matrices can be predicted using (\ref{HAR}). However, this classic approach is inherently unstable due to the following challenges:
\begin{itemize}
	\item[$\bullet$] The received pilot symbols are corrupted by unpredictable noise, resulting in unreliable estimation of AR coefficients $\bm{\gamma}$.
	\item[$\bullet$] The observed CSI is obtained from channel estimation, which is subject to inevitable estimation errors. As a result, the ideal AR relationship in (\ref{HAR}) is often poorly satisfied in practice.
	\item[$\bullet$] To maintain communication efficiency, it is necessary to predict as long CSI as possible based on a limited number of known CSI samples. This requires iterative prediction, during which prediction errors accumulate over time, resulting in progressively degraded accuracy for later CSI predictions.
\end{itemize}

\begin{rmk}
	Even under ideal noise-free conditions with perfect CSI, numerical simulations show that the difference equations derived from the above method are unstable according to the Von Neumann criterion. This lack of numerical stability indicates that the approach is not robust and, therefore, cannot be considered a reliable algorithm. Nonetheless, \textbf{Lemma~\ref{lemma1}} still uncovers an inherent relationship among adjacent channel matrices, which motivates us to adopt a data-driven prediction strategy using deep learning (DL).
\end{rmk}

\subsection{LSTM-based Pilot Prediction Module Design}

Although the model-driven channel predictor derived from Theorem~\ref{theorem1} does not yield accurate or numerically stable channel predictions, we therefore turn to a data-driven DL approach.

\begin{figure*}[t]
	\centering
	\includegraphics[width=0.8\textwidth]{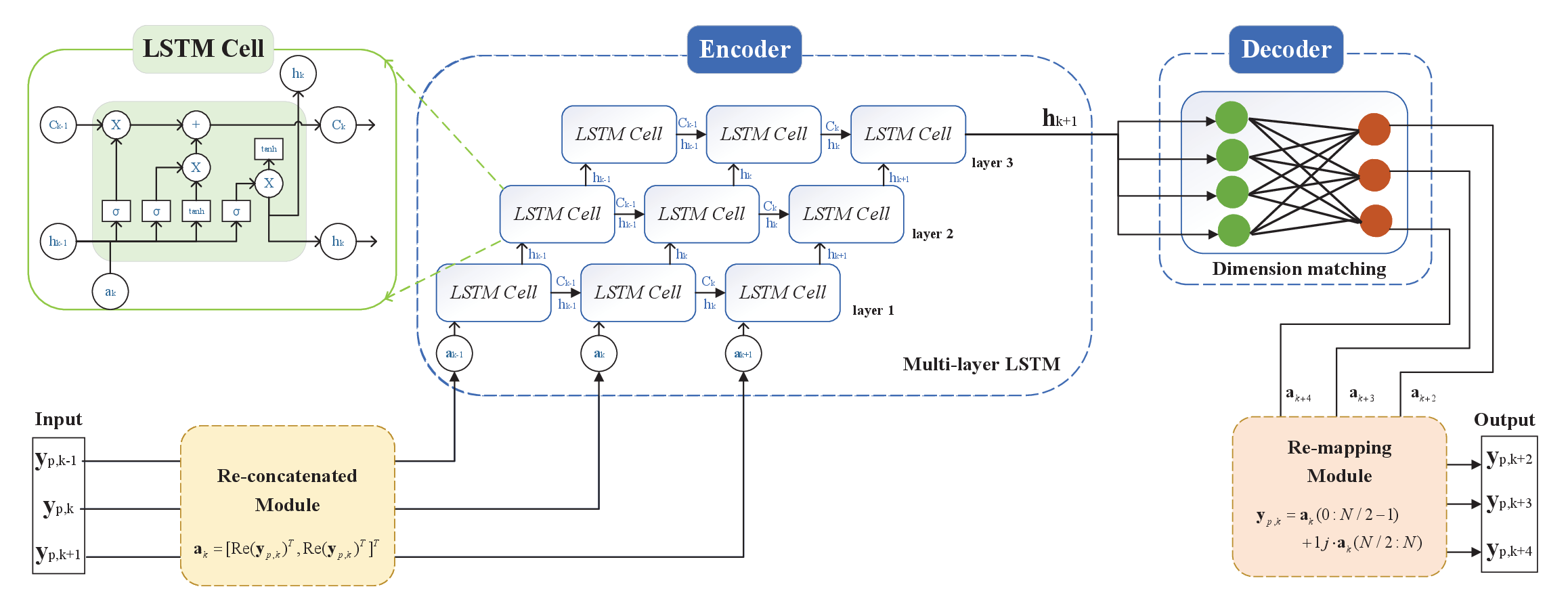}
	\caption{Structure of LSTM-based prediction module.}
	\label{fig:nn}
\end{figure*}

Here, we adopt a pilot symbol prediction scheme based on \textbf{Lemma~\ref{lemma1}}, i.e., the virtual pilot symbol scheme introduced in Section~\ref{II}, instead of directly predicting the channel matrix according to \textbf{Theorem~\ref{theorem1}}. The reason is that symbol-level prediction is affected only by noise, whereas matrix-level prediction is simultaneously impacted by both noise and channel estimation errors. Owing to the strong ability of the LSTM network to capture dependencies among elements in sequence, an LSTM-based architecture is adopted for sequence-to-sequence prediction, as illustrated in Fig.~\ref{fig:nn} (shown at the top of the next page).

To be specific, the dependencies among received pilot symbols are exploited for future symbol prediction, and consists of three key modules: a \textbf{Re-concatenation Module}, a \textbf{Multi-layer LSTM Encoder Module}, and a \textbf{Liner Decoder and Re-mapping Module}.

\subsubsection{Re-concatenation Module}

To adapt complex-valued input data for real-valued NN operations, each pilot symbol is transformed into a real-valued vector by concatenating its real and imaginary parts, i.e.,
\begin{equation}\label{input}
	\mathbf{a}_{k}= [\operatorname{Re}\left(\mathbf{y}_{\text{p},k}\right)^T, \operatorname{Im}\left(\mathbf{y}_{\text{p},k}\right)^T]^T.
\end{equation}
Subsequently, the vector series $\{\mathbf{a}_{k}, \ldots, \mathbf{a}_{k+Q-1}\}$ of length $Q$ is used as the input to the multi-layer LSTM network. Note that the input length $Q$ should satisfy the condition implied by \textbf{Theorem~\ref{theorem1}} as follows:
\begin{corollary} \label{corollary1}
	Within the channel coherence time and bandwidth, the prediction length $Q$ must be greater than or equal to the multi-path number $P$.
\end{corollary}

\subsubsection{Multi-layer LSTM Encoder Module}
The encoder consists of multiple stacked LSTM layers that model the dependencies among pilot symbols. Each LSTM layer contains $Q$ LSTM cells, with each cell containing a hidden state $\mathbf{h}_k$ and a cell state $\mathbf{C}_k$, enabling the network to capture both short-term and long-term correlations. The use of multiple LSTM layers enhances the model’s capability to extract higher-level features and improves prediction accuracy. The hidden state $\mathbf{h}_{k+1}$ of the final LSTM layer summarizes the dependencies of the entire input sequence and serves as the extracted feature vector for subsequent prediction. The corresponding operation can be expressed as
\begin{equation}\label{output}
	\mathbf{h}_{k+1}= \textsl{LSTM}(\mathbf{a}_{k-Q+1},\ldots,\mathbf{a}_{k}) \in \mathbb{C}^{R \times 1},
\end{equation}
where $R$ denotes the dimension of the hidden state, and $\textsl{LSTM}(\cdot)$ represents the nonlinear mapping implemented by the LSTM network. 

\subsubsection{Decoder and Re-mapping Module}
The feature vector $\mathbf{h}_{k+1}$ captures the dependencies among the input symbols, which are then utilized by the decoder to predict future symbols. Specifically, the decoder is implemented as a fully connected network (FCN) that performs prediction based on the extracted feature vector $\mathbf{h}_{k+1}$. The FCN remaps $\mathbf{h}_{k+1}$ to ensure dimensional consistency with the predefined prediction length, thereby generating the subsequent feature sequence. The corresponding process can be formulated as
\begin{equation}\label{remap}
	\{\mathbf{a}_{k+1},\ldots,\mathbf{a}_{k+M}\}= \textsl{FCN}(\mathbf{h}_{k+1}),
\end{equation}
where $M$ denotes the prediction length, and $\textsl{FCN}(\cdot)$ denotes the linear transformation realized by the FCN.

Finally, the prediction result $\{\mathbf{a}_{k+1},\ldots,\mathbf{a}_{k+M}\}$ is reconstructed into complex-valued predicted pilot symbols from the real-valued outputs, which can be expressed as
\begin{equation}\label{reconstruct}
	\mathbf{y}_{\text{p},k}= \mathbf{a}_{k}(1:N)+1j\cdot\mathbf{a}_{k}(N+1:2N).
\end{equation}
\subsubsection{Training Strategy}
The proposed predictor is trained offline, where observations $\mathbf{a}_{k}$ of AF-domain pilot symbols are generated under various channel conditions through simulations to guarantee generalization. The network parameters $\bm{\Theta}$ are optimized using the minimum mean square error (MMSE) criterion, with the loss function is defined as
\begin{equation}\label{loss}
	J(\bm{\Theta})= \mathbb{E}\{\Vert\tilde{\mathbf{a}}_{k} - \mathbf{a}_{k}\Vert_2^2\},
\end{equation}
where $\tilde{\mathbf{a}}_{k}$ denotes the predicted pilot symbol, and $\mathbf{a}_{k}$ denotes the
real pilot symbol.

\section{Numerical Results}

In this section, the effectiveness of the proposed scheme is evaluated via Monte Carlo simulations. {To fit the reality, we utilize the 3GPP NTN time delay line (NTN-TDL) channel model~\cite{3gpp.38.811}, and} the key simulation parameters are summarized in Table~\ref{tab:sim_params}. In addition, each slot consists of 12 OFDM symbols with {a pilot-to-data symbol number ratio of 1:2 (i.e., 4 pilot symbols and 8 data symbols)}, and the simulation is performed over more than 1000 slots. Based on the simulation results reported in~\cite{3gpp.38.821, Kumar2025OJCOMS}, the residual Doppler shift after pre-compensation is approximately 2100~Hz. Accordingly, in this paper, the normalized residual Doppler shift is limited to the range of $[-0.1, 0.1]$. For AF-domain channel estimation, the algorithm presented in~\cite{Tang2025ICCC} is employed. %To assess the prediction performance, we adopt the normalized mean square error (NMSE), which is defined as
%\begin{equation} 
%	\text{NMSE}  = 10\log_{10}\frac{\Vert\tilde{\mathbf{a}}_{k} - \mathbf{a}_{k}\Vert_2^2}{\Vert{\mathbf{a}_{k}}\Vert_2^2}.
%\end{equation}
\begin{table}[t]
	\caption{Simulation Parameters}
	\label{tab:sim_params}
	\centering
	\begin{tabular}{|l|l|}
		\hline
		\textbf{Parameter} & \textbf{Value} \\
		\hline
		Subcarrier number & $N = 64$ \\
		Carrier frequency & $f_c = 2~\text{GHz}$ \\
		Subcarrier spacing & $\Delta f = 30~\text{kHz}$ \\
		Total number of paths & $P = 3$ \\
		% Maximum speed & $v_{\max} = 7.5~\text{km/s}$ \\
		Maximum normalized delay & $l_{\max} = 3$ \\
		Power delay profile & [0 -4.675 -6.482] dB \\
		Fading distribution & Rayleigh \\
		\hline
	\end{tabular}
\end{table}

\begin{figure}
	\centering
	\includegraphics[width=0.5\textwidth]{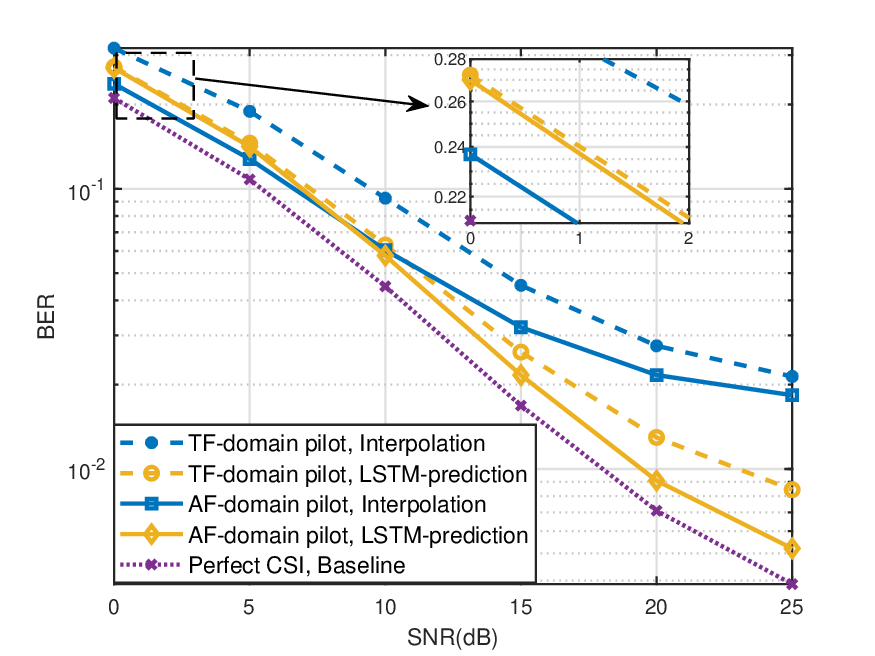}
	\caption{BER performance comparison among different schemes.}\label{fig:1}
\end{figure}

We first evaluate the bit error rate (BER) performance versus the pilot symbol signal-to-noise ratio (SNR), as shown in Fig.~\ref{fig:1}. In this figure, ``TF-domain pilot, Interpolation” represents the conventional OFDM transmission scheme, ``AF-domain pilot, Interpolation” denotes the proposed scheme without prediction, {``TF-domain pilot, LSTM-prediction” represents the OFDM scheme enhanced with an LSTM-based channel predictor, while ``AF-domain pilot, LSTM-prediction” corresponds to the proposed scheme equipped with LSTM-based channel prediction}. The above scheme without predictor employ the CP-based Doppler shift estimation and compensation scheme at the receiver to mitigate the effects of Doppler~\cite{Nishad2013CICT}. As illustrated, the proposed schemes significantly outperform the conventional TF-domain pilot based OFDM transmission method under the same SNR conditions. This improvement arises because the AF-domain channel estimation based on (\ref{AO-relation}) effectively captures the ICI components that are typically neglected in conventional frequency-domain estimators. However, in the high-SNR region, both TF-domain and AF-domain pilot based schemes without channel prediction exhibit an error floor, primarily caused by interpolation errors under high Doppler shifts. In contrast, the LSTM-based predictor eliminates this error floor by leveraging the AR property of the received pilot symbols to achieve more accurate prediction. As the SNR increases, the performance of the proposed prediction scheme approaches that of the “Perfect CSI, Baseline”. {In comparison, the OFDM scheme with predictor deviates more, as the impact of ICI becomes increasingly non-negligible.} It is also observed that at low SNRs, the LSTM-based scheme performs worse than the interpolation-based one. This is mainly because the NN was trained at high SNR, resulting in weaker robustness under noisy conditions. Nevertheless, the overall performance of the proposed scheme remains superior to the conventional OFDM system, and improving its robustness at low SNR will be a key focus in our future work.

\begin{figure}
	\centering
	\includegraphics[width=0.5\textwidth]{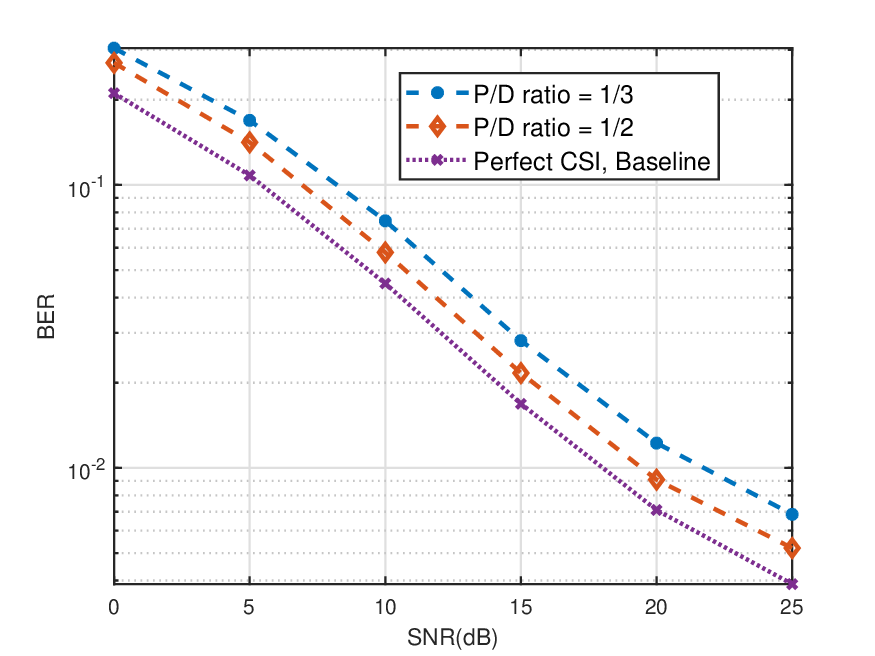}
	\caption{BER performance under different {pilot-to-data symbol number ratios}.}\label{fig:2}
\end{figure}

As a step further, Fig.~\ref{fig:2} illustrates the BER performance under different pilot-to-data symbol number ratios, demonstrating the inherent trade-off between spectral efficiency and BER performance. As observed, schemes with a higher ratio achieve better BER performance compared to those with a lower ratio. This improvement arises from longer input sequences and shorter prediction horizons in virtual pilot and channel prediction, allowing the network to learn more accurate temporal dependencies and enhance prediction accuracy. However, this improvement comes at the cost of reduced spectral efficiency, as more resources are allocated to pilot transmission rather than data.

\section{Conclusion}

In this paper, we proposed an enhanced OFDM transmission scheme for high-mobility LEO satellite communications by incorporating AF-domain processing and LSTM-based channel prediction. To mitigate the severe Doppler effects, we established the relationship between the affine frequency and conventional frequency domains and designed an AF-domain-based pilot based TF-domain channel estimation method. Moreover, by exploiting the AR property among adjacent channels, we proposed an LSTM-based predictor to reduce interpolation errors and alleviate the problem of outdated CSI. Simulation results verified the effectiveness of the proposed framework, demonstrating its strong potential for practical deployment in LEO satellite communication systems.
\bibliographystyle{IEEEtran}%
\bibliography{bib/bibfile}

\vfill

\end{document}